\documentclass[10pt,conference]{IEEEtran}

\usepackage{amsmath,amssymb,amsfonts}
\usepackage{graphicx}
\usepackage{textcomp}
\usepackage{xcolor}
\usepackage[backend=biber,style=ieee,natbib=true,url=false,eprint=false,doi=false,maxnames=2]{biblatex}
\usepackage{hyperref}
\usepackage{booktabs}
\usepackage{tabularx}
\usepackage{array}
\usepackage{makecell}
\usepackage{threeparttable}
\usepackage{float}

\newcolumntype{Y}{>{\raggedright\arraybackslash}X}

\begin{document}

\title{Hybrid quantum-classical end-to-end pipeline for solving MILPs: a vehicle routing case study}

\author{
\IEEEauthorblockN{1\textsuperscript{st} Camille de Valk}
\IEEEauthorblockA{
\textit{Capgemini's Quantum Lab}\\
Utrecht, The Netherlands\\
\textit{Leiden University}\\
Leiden, The Netherlands\\
camille.de.valk@capgemini.com}
\and
\IEEEauthorblockN{2\textsuperscript{nd} Koen Reerink}
\IEEEauthorblockA{
\textit{Capgemini's Quantum Lab}\\
Utrecht, The Netherlands}
\and
\IEEEauthorblockN{3\textsuperscript{rd} Siert Sebus}
\IEEEauthorblockA{
\textit{Capgemini's Quantum Lab}\\
Utrecht, The Netherlands}
\and
\IEEEauthorblockN{4\textsuperscript{th} S\'{e}bastian de Bone}
\IEEEauthorblockA{
\textit{Netherlands Ministry of Defence}\\
The Hague, The Netherlands}
}

\maketitle

\begin{abstract}
We demonstrate an end-to-end hybrid quantum–classical optimisation framework based on Benders decomposition, capable of solving mixed-integer linear programming (MILP) problems. The framework builds on a previously presented hybrid quantum-classical end-to-end pipeline based on Multiple Cuts via Multiple Solutions (MCMS) Benders decomposition where the cut selection step was performed on quantum annealing hardware \cite{paterakis2023hybrid}. We extend this with gate-based QAOA implementations for both tensor network emulators and superconducting quantum hardware. The Vehicle Routing Problem (VRP) is used as a representative case study and we run the pipeline end-to-end on 10 permutations of a standardised benchmarking instance ($20$ customers and $4$ vehicles from QOptLib \cite{osaba2023qoptlib}) with a classical solver performing the cut selection step. We find that for our instances, only a small fraction of the compute in classical MCMS Benders decomposition is spent on the cut selection step. For a full hybrid end-to-end assessment, we run the pipeline for a toy problem with MPS-JuliQAOA, a powerful tensor network emulator, to execute QAOA. Here, the majority of the time is spent on the cut selection step, deeming quantum advantage of this framework unlikely at problems of this size. This highlights the need for more large-scale benchmarking research when more powerful (QPU) QUBO solvers are available.
\end{abstract}


\section{Introduction}
Recent advances in quantum computing have renewed interest in its potential to address computationally intractable problems \cite{phillipson2024quantum}. The development of quantum algorithms, such as the Quantum Approximate Optimization Algorithm (QAOA) \cite{farhi2014quantum}, along with steady improvements in gate-based hardware platforms, has expanded the range of applications that can be explored on near-term quantum devices \cite{wendin2017quantum}. Although fault-tolerant quantum computing remains a long-term goal, current Noisy Intermediate-Scale Quantum (NISQ) systems already enable experimentation with hybrid quantum–classical workflows, where quantum routines are embedded within classical optimisation pipelines \cite{mcclean2016theory, jiang2025advancements}. These developments have motivated both academia and industry to investigate practical use cases in which quantum computing may provide value beyond purely theoretical demonstrations.

Across many industries, high-impact decision-making tasks can be formulated as linear optimization problems \cite{golden2024unexpected}.
In practice, such problems often are Mixed-Integer Linear Programs (MILPs). These are linear problems containing both discrete and continuous variables. MILPs are computationally challenging to solve at larger scales due to their combinatorial nature, and even state-of-the-art classical solvers may struggle to find optimal solutions within acceptable timeframes for large instances \cite{achterberg2013mixed}. This creates a strong incentive to explore alternative computational paradigms, including hybrid quantum–classical approaches, that can accelerate or enhance existing solution methods.

To address the complexity of MILPs, decomposition techniques such as Benders decomposition (BD) \cite{Bender1962partitioning} have been widely studied, also in combination with quantum optimisation \cite{paterakis2023hybrid, zhao2022hybrid, naghmouchi2024mixed, ellinas2024hybrid, yoshihara2026accelerating, lopez2026performance}. Benders decomposition separates a large optimisation problem into a master problem and one or more subproblems, which are solved iteratively while exchanging information in the form of constraints, or “cuts”. This approach is particularly attractive for large-scale applications due to its flexibility and ability to provide intermediate feasible solutions. However, a well-known limitation of Benders decomposition is its often slow convergence, which depends critically on the selection of informative cuts \cite{rahmaniani2017benders}.

\citeauthor{paterakis2023hybrid} has explored heuristic and quantum-inspired approaches to improve this cut selection process by formulating it as a combinatorial optimisation problem amenable to quantum solving techniques \cite{paterakis2023hybrid}, quantum annealing specifically. The method formulates the cut selection step as a minimum set cover problem, which is translated to a Quadratic Unconstrained Binary Optimisation (QUBO) problem. While the approach is able to run on a small toy problem, it is clear that the translation to QUBO and minor embedding yields significant overhead when scaling the problem \cite{paterakis2023hybrid}. This is one of the challenges for quantum annealing systems outlined in \cite{quinton_quantum_2025}. The challenges of annealing motivate the question whether other quantum approaches can be used.

In this paper, we present an end-to-end hybrid quantum–classical optimisation framework that integrates QAOA subroutines into a classical Benders decomposition pipeline. Our method solves the cut selection step using QAOA, a gate-based quantum computing approach, enabling execution on both tensor network emulators \cite{fermioniq, fermioniq-ava} and superconducting quantum hardware \cite{qiskit2024, jiang2025advancements}. We test the feasibility and the performance of our framework on a Vehicle Routing Problem (VRP) instance. Such an instance can be formulated as a MILP and thus be solved with our quantum-enhanced Benders decomposition algorithm. The VRP represents a canonical logistics optimisation problem with direct relevance to real-world applications such as supply chain management and transportation planning \cite{mor2022vehicle}. As such, this end-to-end setup is a practical and reproducible step toward quantum-enhanced large-scale optimisation in real-world settings. Beyond algorithmic performance, the study provides empirical insights into where quantum methods can add value within full optimisation pipelines, while also highlighting current limitations and future opportunities for quantum advantage in the context of optimisation.

\section{Preliminaries}

\subsection{Benders Decomposition}
Benders decomposition (BD) \cite{Bender1962partitioning, rahmaniani2017benders} is a decomposition technique that exploits the separable structure of problems with both continuous and integer variables. A MILP can be written as
\begin{equation}
\min_{\mathbf{x},\,\mathbf{y}} \ \mathbf{c}^T \mathbf{x} + \mathbf{d}^T \mathbf{y}
\quad \text{s.t.} \quad
\mathbf{A}\mathbf{x} + \mathbf{B}\mathbf{y} \ge \mathbf{b}, \quad
\mathbf{x} \ge 0, \quad
\mathbf{y} \in Y,
\end{equation}
where $\mathbf{x}$ are continuous variables and $\mathbf{y}$ are integer (binary) variables. The vectors $\mathbf{b}$, $\mathbf{c}$, and $\mathbf{d}$, and the matrices $\mathbf{A}$ and $\mathbf{B}$ contain real-valued entries that define the objective function and constraints of the problem. $Y$ represents a set of integers under constraints.

BD separates the original problem into a master problem (MP), defined over the integer variables $\mathbf{y}$, and a subproblem (SP), defined over the continuous variables $\mathbf{x}$. For a fixed assignment $\hat{\mathbf{y}}$, the subproblem becomes
\begin{equation}
\text{SP}(\hat{\mathbf{y}}): \quad
\min_{\mathbf{x}} \ \mathbf{c}^T \mathbf{x}
\quad \text{s.t.} \quad
\mathbf{A}\mathbf{x} \ge \mathbf{b} - \mathbf{B}\hat{\mathbf{y}}, \quad
\mathbf{x} \ge 0.
\end{equation}

The MP contains only the integer variables $\mathbf{y}$ and an auxiliary variable $z$. This auxiliary variable represents the contribution of the omitted continuous
variables to the objective. The MP starts unconstrained. As Benders cuts are added, it iteratively approximates the original MILP:
\begin{equation}
\label{eq:bd-mp}
\text{MP:} \quad
\min_{\mathbf{y},\, z} \ z
\quad \text{s.t.} \quad
\{\text{cuts}\}, \quad
\mathbf{y} \in Y.
\end{equation}

At each iteration, the MP is solved, producing a candidate solution $\hat{\mathbf{y}}$, from which a subproblem is created. Based on the solution to the dual of the SP, either a feasibility cut or an optimality cut is generated and added to the set of cuts, denoted $\{\text{cuts}\}$ in~\eqref{eq:bd-mp}. This iterative process refines upper and lower bounds until convergence \cite{rahmaniani2017benders}.

To accelerate convergence, a multi-cut variant of BD can be employed, where multiple candidate solutions $\{\hat{\mathbf{y}}_i\}$ are generated per iteration and multiple subproblems are solved in parallel \cite{paterakis2023hybrid, beheshti2019accelerating}. This produces multiple cuts per iteration, significantly improving convergence speed \cite{beheshti2019accelerating}. However, adding all generated cuts can lead to an overly large and computationally expensive master problem. One way of minimizing the risk of adding inefficient or redundant cuts to the MP is to perform heuristics-based cut selection every iteration \cite{rahmaniani2017benders, beheshti2019accelerating}. Depending on the chosen heuristic, finding optimal cuts can be an NP-hard problem by itself. Our method employs quantum optimisation to perform this step. This is explained further in Section \ref{sec:methodology} and Figure \ref{fig:framework-overview}.

\subsection{Vehicle Routing Problems}
The Vehicle Routing Problem (VRP) is a widely studied combinatorial optimisation problem that arises naturally in logistics domains such as supply chain management and transportation planning \cite{mor2022vehicle}. In this setting, a fleet of vehicles must service a set of distributed customers from a central depot. Each customer is associated with a demand, while each vehicle is subject to capacity constraints. The objective is to construct routes that serve all customers while minimizing total travel cost, subject to feasibility constraints on capacity. Due to the combinatorial nature of route assignment and sequencing, the problem quickly becomes intractable for classical exact methods as instance size increases.

We consider the capacitated VRP defined on a directed graph $G = (V, E)$, where the node set $V$ consists of a single depot $0$ and a set of customers $C = \{1, \dots, n\}$ connected by edges $E$. A fleet of $k$ homogeneous vehicles with identical capacity $Q$ is available. Each customer $i \in C$ has a demand $d_i$. Travel between nodes $(i, j) \in E$ incurs a cost $c_{ij}$. The objective is to construct a set of routes such that all customers are visited exactly once and total travel cost is minimized.

We adopt a MILP formulation for the VRP \cite{miller1960integer}. We define binary decision variables:
\[
x_{ij} =
\begin{cases}
1 & \text{if a vehicle travels directly from node } i \text{ to node } j, \\
0 & \text{otherwise}.
\end{cases}
\]
Additionally, let $u_i\in \mathbb{R}$ denote auxiliary variables for subtour elimination and load tracking. The MILP formulation is:
\begin{align}
\label{eq:objective}
\min \sum_{i \in V} \sum_{\substack{j \in V \\ j \neq i}} &c_{ij} x_{ij}\\
\text{s.t.} \nonumber\\
\label{eq:outgoing}
\sum_{\substack{j \in V \\ j \neq i}} x_{ij} &= 1 \quad \forall i \in C\\
\label{eq:incoming}
\sum_{\substack{i \in V \\ i \neq j}} x_{ij} &= 1 \quad \forall j \in C\\
\label{eq:depot}
\sum_{j \in C} x_{0j} = k, &\quad \sum_{i \in C} x_{i0} = k\\
\label{eq:mtz}
u_i - u_j + Q x_{ij}\leq Q - d_j \quad &\forall i \neq j,\ i,j \in C\\
\label{eq:capacity_bounds}
d_i \leq u_i \leq Q \quad& \forall i \in C\\
\label{eq:domains}
x_{ij} \in \{0,1\},& \quad u_i \geq 0
\end{align}
The objective function~\eqref{eq:objective} minimizes the total travel cost. Constraints~\eqref{eq:outgoing} and~\eqref{eq:incoming} ensure that each customer is visited exactly once, with one incoming and one outgoing edge. The depot constraints~\eqref{eq:depot} enforce that exactly $k$ vehicles depart from and return to the depot. The Miller-Tucker-Zemlin \cite{miller1960integer} constraints~\eqref{eq:mtz}, together with bounds~\eqref{eq:capacity_bounds}, eliminate subtours and ensure that vehicle capacity is not exceeded along any route. Finally,~\eqref{eq:domains} defines the binary routing decisions and non-negative load variables.

\section{Methodology}
\label{sec:methodology}

\begin{figure*}[!t]
    \centering
    \includegraphics[width=0.71\linewidth]{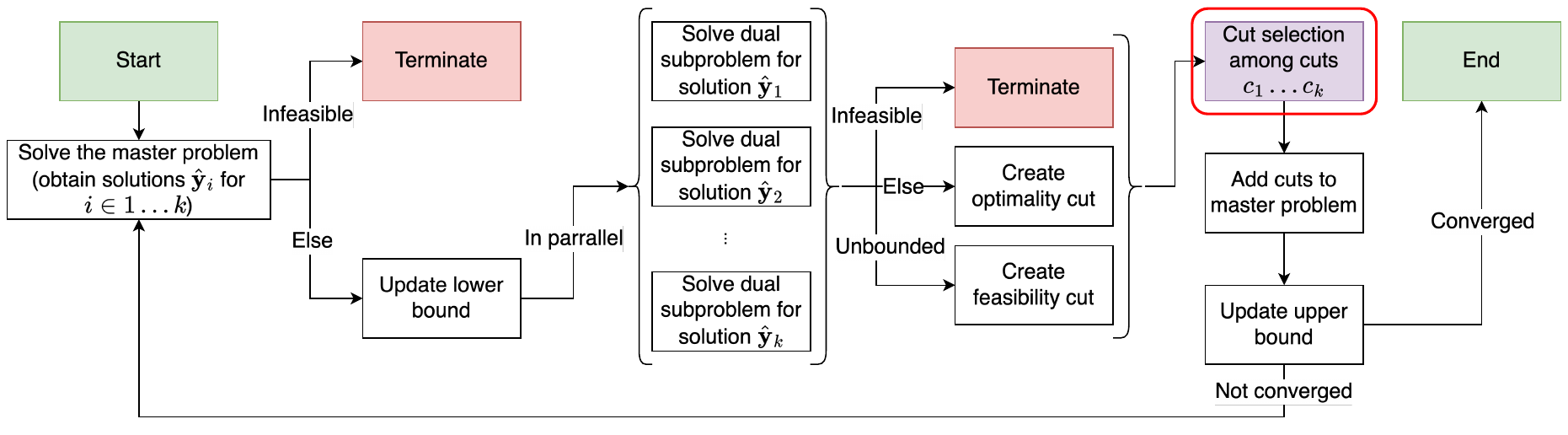}
    \caption{A visual overview of our framework, which is adapted from \citeauthor{paterakis2023hybrid} \cite{paterakis2023hybrid}. It employs the HQC-MCMS Benders decomposition algorithm, delegating the cut selection step (denoted in purple) to the QAOA algorithm.}
    \label{fig:framework-overview}
\end{figure*}

Our framework builds on the Hybrid Quantum-Classical Multiple Cuts via Multiple Solutions (HQC-MCMS) approach introduced by \citeauthor{paterakis2023hybrid} \cite{paterakis2023hybrid}, in which the cut selection step of MCMS Benders decomposition is solved as a QUBO problem (Section \ref{subsec:cut-selection-set-cover}) using quantum annealing. We extend this framework to gate-based quantum hardware by solving the cut selection QUBO with QAOA (Section~\ref{subsec:set-cover-as-qubo}), enabling execution on both tensor network emulators and superconducting quantum processors (Section \ref{subsec:solving-qubo}). A visual overview of our method is given in Figure \ref{fig:framework-overview}.

\subsection{Cut Selection via Set Cover}
\label{subsec:cut-selection-set-cover}

To control the growth of the master problem, we use a cut selection procedure that selects a subset of informative cuts from the candidate pool. We formulate this as a minimum set cover problem, where each cut corresponds to a set covering certain infeasible solutions or regions of the search space.

Let $\mathcal{C} = \{c_1, \dots, c_m\}$ denote the set of candidate cuts and $\mathcal{E} = \{e_1, \dots, e_n\}$ the set of elements to be covered (e.g., infeasible master solutions). We construct a binary matrix $M \in \{0,1\}^{m \times n}$, where $M_{ij} = 1$ if cut $c_i$ eliminates or covers element $e_j$.

The cut selection problem is then formulated as:
\begin{equation}
\min_{\mathbf{x}} \sum_{i=1}^m x_i
\end{equation}
\begin{equation}
\text{s.t.} \quad \sum_{i=1}^m M_{ij} x_i \ge 1 \quad \forall j = 1, \dots, n,
\end{equation}
\begin{equation}
x_i \in \{0,1\} \quad \forall i = 1, \dots, m.
\end{equation}

This formulation ensures that all relevant elements are covered while minimizing the number of selected cuts. By selecting only the most informative cuts, we reduce redundancy and limit the growth of the master problem, improving overall computational efficiency.

\subsection{Set Cover as QUBO for Quantum Computers}
\label{subsec:set-cover-as-qubo}
The cut selection problem is inherently combinatorial and can be naturally expressed using binary decision variables indicating whether a cut is selected. We therefore follow \cite{paterakis2023hybrid} and reformulate this problem as a Quadratic Unconstrained Binary Optimisation (QUBO) instance, allowing both the objective and coverage constraints to be encoded into a single cost function. Unlike \cite{paterakis2023hybrid}, who employ quantum annealing for QUBO optimization, we adopt the gate-based Quantum Approximate Optimization Algorithm (QAOA) \cite{farhi2014quantum}. Since quantum annealing typically requires a costly minor-embedding procedure \cite{quinton_quantum_2025}, QAOA represents a promising alternative, as it is extensively studied for QUBO optimization on gate-based quantum devices \cite{blekos_review_2024} and does not require minor embedding. QAOA operates in a hybrid quantum--classical loop, where a parameterised quantum circuit explores the solution space and a classical optimiser updates the circuit parameters. Within our framework, we use Qiskit \cite{qiskit2024} to construct the QUBO from the minimum set cover of candidate cuts at each Benders iteration. It can be solved using QAOA on either emulators or quantum hardware. The resulting solution identifies a subset of cuts to be added to the master problem. Note that QUBOs are by definition unconstrained, so the resulting selection might not form a (minimum) set cover. Since QAOA is stochastic and we use multiple shots (see Table \ref{tab:qaoa-configurations}), we can filter the lowest-cost sample of the QAOA that indeed forms a set cover.

\subsection{Solving QUBOs in our end-to-end pipeline}
\label{subsec:solving-qubo}
A large contribution of our work is the interfacing between Hybrid Quantum-Classical Multiple Cuts via Multiple Solutions (HQC-MCMS) Benders decomposition, proposed by \citeauthor{paterakis2023hybrid}, and other QUBO solvers. We integrated three QAOA QUBO solvers. All can run the end-to-end pipeline as outlined in Figure \ref{fig:framework-overview}.
\subsubsection{Fermioniq’s Ava \cite{fermioniq, fermioniq-ava}}
Fermioniq's Ava tensor network circuit emulator can accurately emulate (noisy) quantum circuits  \cite{fermioniq, fermioniq-ava}, which we used to emulate QAOA quantum circuits. To run QAOA on their emulators, we use the VQE functionality, where we specify the ansatz as the canonical Ising QUBO cost operator and mixer operator. Using VQE this way is equivalent to using QAOA \cite{farhi2014quantum}.

\subsubsection{MPS-JuliQAOA \cite{mps-juliqaoa}}
MPS-JuliQAOA is an open-source QAOA emulator built in Julia using ITensor for Matrix Product State (MPS) emulation \cite{mps-juliqaoa}. It can be used to directly pass an Ising Hamiltonian, which is then optimised by emulating QAOA. For us, the main difference with Fermioniq's emulator is that MPS-JuliQAOA is open-source, so we can run it on any hardware, not limited to hardware where Fermioniq's emulator is available.

\subsubsection{IBM Quantum via Qiskit \cite{qiskit2024}}
Qiskit can be used to directly interface with QPUs. Our pipeline currently only connects to IBM Quantum QPUs, but with little effort, it can be connected to any QPU available from Qiskit. We note that the connection to QPUs is mainly built for demonstration purposes (see results in \ref{subsec:end-to-end-toy}), as today's QPUs can still be emulated \cite{TN-emulation-of-IBMQ} and \citeauthor{paterakis2023hybrid} already indicated that cut selection in HQC-MCMS Benders decomposition is mainly interesting for large-scale instances beyond the capabilities of today's QPUs \cite{paterakis2023hybrid}. Therefore, running large hardware experiments would be an uninformative endeavour.

\section{End-To-End Case Study}
\label{sec:end-to-end-case-study}
To evaluate the practical applicability of our hybrid quantum-classical framework, we consider an end-to-end case study based on the VRP. We run three experiments, with setups outlined in Tables \ref{tab:mcms-experiments} and \ref{tab:qaoa-configurations}. First, we test the end-to-end pipeline on one benchmarking instance of VRP from the Quantum Optimization Benchmarking Library (QOptLib) \cite{osaba2023qoptlib} to show the entire MCMS Benders decomposition. QOptLib offers standardised, reproducible benchmark instances designed to assess optimisation methods on practically relevant and computationally challenging problems. We consider an instance with $n=20$ customers and $k=4$ vehicles on a fully connected graph $G$. In the full pipeline, the master problem and the minimum set cover are solved using Cbc (Coin-or branch and cut) \cite{forrest2026cbc}. The dual subproblems are solved with HiGHS \cite{highs}, because it can return extreme rays from solutions, necessary for adding feasibility cuts. The results are presented and discussed in Section \ref{subsec:end-to-end-mcms}.

To demonstrate our pipeline indeed runs end-to-end with the proposed QAOA step, we solve a random small toy problem with $n=5$ customers with random demands $d \sim \mathcal{U}(1,7)$ placed randomly on a grid $(x_i, y_i) \overset{\text{i.i.d.}}{\sim} \mathrm{Unif}\bigl(\{0,\dots,9\}^2\bigr)$, $k=3$ vehicles with capacity $Q=8$, and $1$ depot. The costs $c_{i,j}$ are given by the Euclidean distance. Section \ref{subsec:end-to-end-toy} shows and discusses the results.

Finally, in Section \ref{subsec:msc-with-qaoa} we demonstrate the implementation of the different QAOA solvers (Fermioniq, MPS-JuliQAOA, and IBM Quantum via Qiskit) for solving a minimum set cover from a HQC-MCMS Benders iteration with $5$ subproblems. The set cover problem is depicted in the inset of Figure \ref{fig:set-cover-qaoa}.

\begin{table*}[t]
\centering
\caption{Experimental setup for MCMS-based experiments.}
\label{tab:mcms-experiments}
\begin{tabular}{lcc}
\toprule
 & \textbf{QOptLib MCMS} & \textbf{Toy HQC-MCMS} \\
\midrule
Problem type
  & Benchmark VRP \cite{osaba2023qoptlib}
  & Random toy instances \\
Number of instances
  & 1 benchmark, 10 permutations
  & 5 randomly generated (Sec. \ref{sec:end-to-end-case-study}) \\
Master problem (MP) solver
  & Cbc \cite{forrest2026cbc}
  & Cbc \\
Dual subproblem (DSP) solver
  & HiGHS \cite{highs}
  & HiGHS \\
Cut selection solver
  & Cbc
  & MPS-JuliQAOA (Tab. \ref{tab:qaoa-configurations}) with $p=1$ \\
Subproblems per iteration
  & 2, 4, 16, 64
  & 5 \\
Time limit
  & 1.5 hours
  & 4 hours \\
Iteration limit
  & 2500
  & 1000 \\
\bottomrule
\end{tabular}
\end{table*}

\begin{table}[t]
\caption{QAOA solver configurations. For the tensor network emulators, the bond dimension is written as $\chi$.}
\label{tab:qaoa-configurations}
\centering
\footnotesize
\setlength{\tabcolsep}{4pt}
\renewcommand{\arraystretch}{1.1}

\begin{threeparttable}
\begin{tabularx}{\columnwidth}{@{}Y Y c Y@{}}
\toprule
\textbf{Backend} & \textbf{Solver type} & \makecell{$\chi$} & \textbf{Hardware} \\
\midrule
\textbf{MPS-JuliQAOA} \cite{mps-juliqaoa}
& MPS emulator
& 64
& AMD EPYC 9554P \\
\textbf{Fermioniq Ava} \cite{fermioniq, fermioniq-ava}
& DMRG emulator
& 64
& \texttt{cpu-16} \\
\textbf{IBM Quantum} \cite{qiskit2024}
& Superconducting quantum hardware
& n/a
& \texttt{ibm\_strasbourg} \\
\bottomrule
\end{tabularx}

\begin{tablenotes}[flushleft]
\footnotesize
\item All configurations used 1000 shots, QAOA depths $p=\{1,3\}$, and COBYLA \cite{NLopt} with $\rho_{\text{beg}}=\pi/2$ and a maximum of 1000 iterations for classical optimisation of the QAOA parameters.
\end{tablenotes}
\end{threeparttable}
\end{table}

\section{Results}

\subsection{MCMS-Benders decomposition for Benchmarking VRP}
\label{subsec:end-to-end-mcms}

\begin{figure*}[t]
    \centering
    \includegraphics[width=\linewidth]{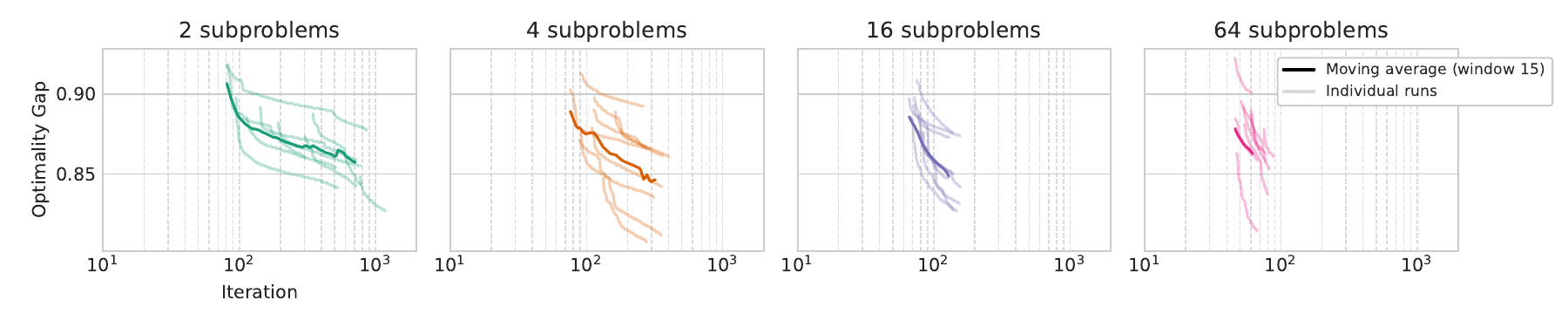}
    \caption{Optimality gap, see Equation \eqref{eq:gap}, over iterations for 10 permutations of one VRP from QOptLib \cite{osaba2023qoptlib}. The gap decreases as the algorithm converges, with different subproblem configurations affecting convergence speed. The more subproblems that are considered, the fewer iterations before the first feasible solution is found.}
    \label{fig:gap}
\end{figure*}

The first result is shown in Figure \ref{fig:gap}. In the figure, we plot the optimality gap defined as
\begin{equation}
\label{eq:gap}
    Gap = \begin{cases}
        \frac{UB - LB}{UB}  & \text{if } 0<UB<\infty, 0\leq LB < \infty \\
        \infty & \text{else}.
    \end{cases}
\end{equation}
Because there was a limit on total time (see Table \ref{tab:mcms-experiments}), not all experiments ran for the same number of iterations. Clearly, the runs with more subproblems had slower iterations and timed out earlier. The moving average with a window of 15 is plotted for iterations with 5 or more runs at that iteration with a solid line. Shaded lines represent individual runs.

We can observe that the end-to-end hybrid pipeline indeed finds feasible solutions and the optimality gap becomes smaller over time. All optimality gaps remained high above a convergence threshold, which is typically below 1\%. From the figures (with a logarithmic $x$-axis) it becomes clear that considering more subproblems leads to feasible solutions in fewer iterations, indicating the value of the MCMS pipeline. Convergence also seems to be faster, though with so few runs, we cannot state that conclusively. The insight that considering more subproblems leads to feasible solutions faster is in line with \citeauthor{paterakis2023hybrid} \cite{paterakis2023hybrid} and \citeauthor{beheshti2019accelerating} \cite{beheshti2019accelerating}. Both highlight the fact, however, that considering more subproblems also adds a computational overhead in the cut selection step.

\begin{figure*}[!t]
    \centering
    \includegraphics[width=\linewidth]{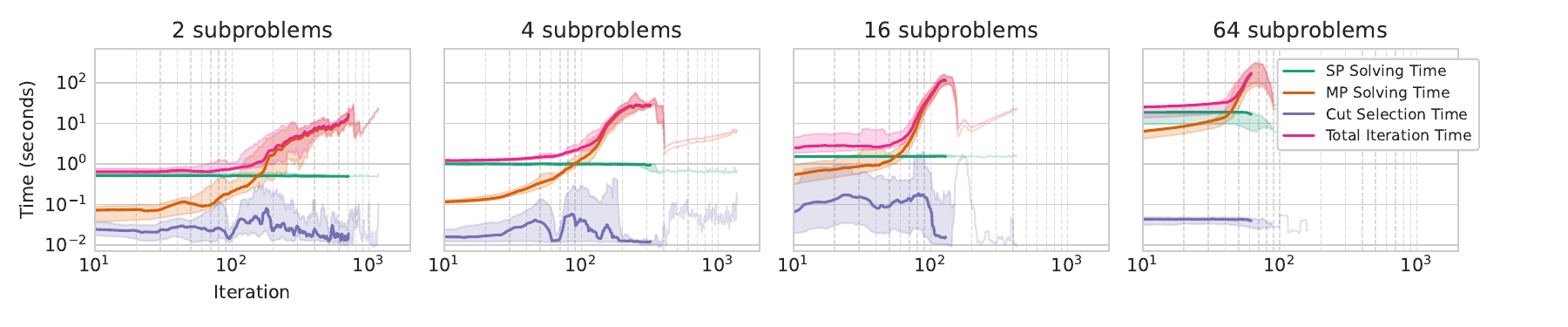}
    \caption{Timings of different parts of the end-to-end optimisation pipeline over iterations for 10 permutations of one VRP from QOptLib \cite{osaba2023qoptlib}. The solid line shows the median and the shaded regions show the 25$^\text{th}$ and 75$^\text{th}$ percentiles. To avoid a noisy line, the median is only plotted when there are 5 or more runs at that iteration. Over time MP problem solving becomes increasingly harder, while SP solving remains roughly the same. Cut selection contributes only to a small part of the iteration time.}
    \label{fig:timing-in-iterations}
\end{figure*}

In Figure \ref{fig:timing-in-iterations}, we show the median time the algorithm spent in different parts of the end-to-end pipeline (MP solving, SP solving, cut selection with minimum set cover). Here, we plot a moving median as a solid line, since the average is easily skewed if there are outliers in this logarithmic range. From the figure, it becomes clear that the majority of the time in an iteration is spent on MP solving. SP solving times remain constant throughout the process, which is to be expected. The time spent on cut selection is orders of magnitude lower than the other parts of the algorithm, which weakens the claim that the cut selection step is a computational bottleneck one should try to solve with quantum computing.

\subsection{End-to-end pipeline for Toy Problem}
\label{subsec:end-to-end-toy}
\begin{figure}
    \centering
    \includegraphics[width=\linewidth]{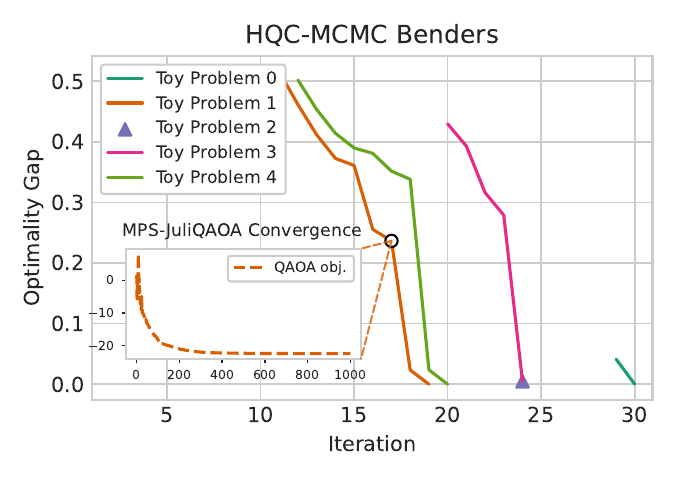}
    \caption{Convergence of our HQC-MCMS Benders solver for 5 random toy problems (Section \ref{sec:end-to-end-case-study}) with MPS-JuliQAOA backend with QAOA depth $p=1$ (Table \ref{tab:qaoa-configurations}). As an example, for toy problem 1 the QAOA convergence trace of the cut selection process (minimum set cover as QUBO with MPS-JuliQAOA) of iteration 17 is plotted in the inset. For problem 3, the solution directly went from unbounded to optimal, indicated by the marker instead of the convergence trace. The wall-clock times for the entire pipeline were $2.9\cdot10^3 \pm 0.8\cdot10^3$ seconds. On average, $99.3\%$ of this time was spent on the cut selection step. One can see that minimum set cover for cut selection can be solved much faster using Cbc in Figure \ref{fig:timing-in-iterations}.}
    \label{fig:toy-convergence}
\end{figure}

In Figure \ref{fig:toy-convergence}, we show the convergence of our HQC-MCMS Benders end-to-end pipeline for 5 toy problems as defined in Section \ref{sec:end-to-end-case-study}. In every iteration, QAOA is solved with MPS-JuliQAOA for the set cover problem, indicated by the inset. We should note the total runtimes of these HQC-MCMS runs are much higher than those of the non-quantum end-to-end pipeline for the benchmarking instance (Figures \ref{fig:gap} and \ref{fig:timing-in-iterations}). This indicates that MPS-JuliQAOA is much worse at solving minimum set cover than Cbc at this scale.

\subsection{Minimum Set Cover with QAOA}
\label{subsec:msc-with-qaoa}
\begin{figure}
    \centering
    \includegraphics[width=\linewidth]{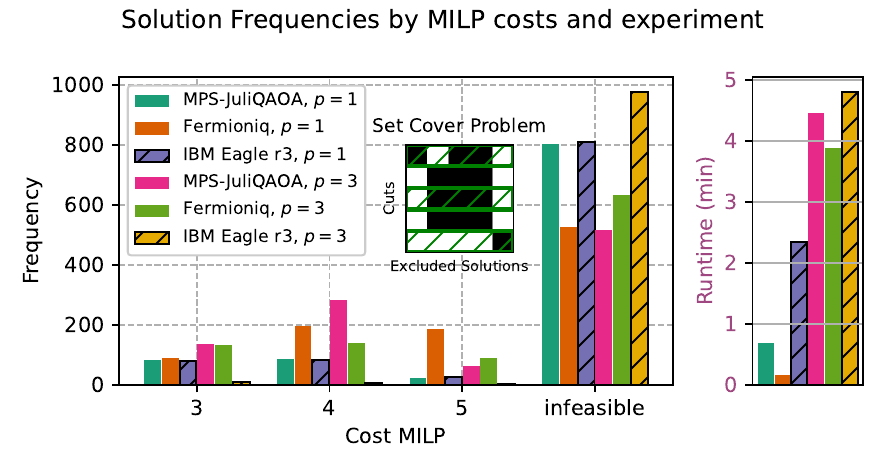}
    \caption{Examples of sample frequencies from a QAOA circuit for different QAOA configurations and their respective QAOA optimisation runtimes (Table \ref{tab:qaoa-configurations}). Circuits were transpiled at Qiskit's highest optimization level (level 3), combining heuristic-optimized layout/routing, KAK-based resynthesis of two-qubit blocks, and commutative gate cancellation. Two additional passes reduce circuit overhead by reordering measurements to avoid SWAPs and removing diagonal gates that don't affect measurement outcomes. No error mitigation techniques were employed. The hardware runs are indicated with a hatch. The set cover problem is shown in the inset, where the cuts coming from the dual subproblems are depicted as rows from the matrix with a black square if they exclude a solution of the master problem. A valid set cover chooses rows such that all the columns are covered. The set cover, shown in green, consists 3 cuts covering all solutions.}
    \label{fig:set-cover-qaoa}
\end{figure}

Figure \ref{fig:set-cover-qaoa} shows the frequencies for the different costs sampled by the backends specified in Table \ref{tab:qaoa-configurations}. It also shows the runtimes for the QAOA. From the frequencies, we can observe that most of the samples yield infeasible solutions, i.e., they are not a valid set cover. Since QAOA solves an unconstrained problem by nature and it samples from a quantum circuit, there may be samples that are not a valid set cover and therefore infeasible. For all configurations, this is the majority of samples. For the emulated QAOA configurations, the QAOA with depth $p=3$ had more samples with optimal cost than for $p=1$, indicating the value of a deeper QAOA circuit. We note that this figure is to demonstrate we ran QAOA on different backends and not to benchmark the performance of these backends or the algorithm. For an extensive benchmark and review on QAOA, we refer to \cite{bucher_towards_2024} and \cite{blekos_review_2024} respectively.

\section{Discussion \& Future Outlook}
This work built a end-to-end MCMS Benders decomposition pipeline and demonstrated that it can run for benchmarking instances in a fully-classical setting (Figures \ref{fig:gap} and \ref{fig:timing-in-iterations}) and for toy instances in a quantum-inspired setting (Figure \ref{fig:toy-convergence}). The cut selection step, framed as minimum set cover, can be solved with different quantum solvers (Figure \ref{fig:set-cover-qaoa}). The quantum step attempts to solve an NP-hard problem, but at the scales considered, the classical method Cbc vastly outperforms any of our QAOA attempts. Furthermore, the classical step considered was the least significant in terms of runtime impact in our setup, highlighting a serious weakness in the argument for using quantum computing at the cut selection step.

From Figures \ref{fig:gap} and \ref{fig:timing-in-iterations}, we can see that the pipeline indeed finds lower and upper bounds (presented as a gap). However, it does not find the optimum within the time limit (1.5 hours). We also note that MP solving time increases with the number of subproblems, indicated by the point at which the orange lines start in the different subfigures of Figure \ref{fig:timing-in-iterations}. This is because we extract $n$ feasible solutions in a sequential manner. Powerful commercial solvers can extract multiple feasible solutions at much lower computational cost, which would greatly reduce the iteration time for settings with many subproblems.

In future research, the scaling of the (HQC)-MCMS Benders decomposition should be explored to conclusively confirm or rule out the significance of cut selection in MCMS Benders decomposition in terms of timing. We note that our research worked with open-source MILP solvers and that using better solvers could also change the view on the potential of this algorithm. Another weakness in our demonstrated approach is the conversion of minimum set cover to QUBO, also indicated by \cite{paterakis2023hybrid}, because it adds a substantial overhead. Other attempts at speeding up Benders decomposition could be considered, for example by framing the cut selection step as a maximum independent set problem, well-suited for neutral atoms quantum computing hardware \cite{ebadi_quantum_2022}, or as a Higher-Order Unconstrained Binary Optimisation (HUBO) problem, which can be solved on a gate-based QPU with tailored hybrid quantum algorithms like Digitized Counterdiabatic Quantum Optimisation Algorithm \cite{romero_bias-field_2025}.
We conclude our discussion by stating that none of these new attempts at cut selection have a serious chance if the cut selection remains such an insignificant part of the MCMS Benders decomposition at scale.

\section{Conclusion}
\label{sec:conclusion}
This paper extended the Hybrid Quantum-Classical Multiple Cuts via Multiple Solutions (HQC-MCMS) Benders decomposition, presented by \citeauthor{paterakis2023hybrid} in \cite{paterakis2023hybrid} to gate-based quantum computing and tested it on a benchmarking instance of an industry-relevant case study: the Vehicle Routing Problem. We demonstrated the pipeline can work for benchmarking instances of the VRP and that the pipeline works end-to-end with a quantum step using QAOA on a toy instance. Our results show that the pipeline can run both in a fully classical setting and in a quantum-inspired setting, though we find that in our settings, the fully classical pipeline was much stronger.

\section*{Acknowledgment}
This work was performed using the ALICE compute resources provided by Leiden University and \texttt{ibm\_strasbourg}, which is one of the IBM Quantum Eagle processors. The code used for the experiments and the figures can be found in \cite{devalk_hqc_mcms_benders}.

\printbibliography

\end{document}